\documentclass[fleqn,usenatbib]{mnras}

\usepackage{newtxtext,newtxmath}
\usepackage[T1]{fontenc}

\DeclareRobustCommand{\VAN}[3]{#2}
\let\VANthebibliography\thebibliography
\def\thebibliography{\DeclareRobustCommand{\VAN}[3]{##3}\VANthebibliography}

\usepackage{graphicx}	% Including figure files
\usepackage{amsmath}	% Advanced maths commands

\title[MHD wind-driven discs]{Analytical solutions for the evolution of MHD wind-driven accretion discs}

\author[Shadmehri \& Khajenabi]{
Mohsen Shadmehri$^{1}$\thanks{E-mail: mmshadmehri@gmail.com (MS)}
,
Fazeleh Khajenabi$^{1}$
\\
$^{1}$Department of Physics, Faculty of Sciences, Golestan University, Gorgan 49138-15739, Iran
}

\date{Accepted XXX. Received YYY; in original form ZZZ}

\begin{document}
\label{firstpage}
\pagerange{\pageref{firstpage}--\pageref{lastpage}}
\maketitle

\begin{abstract}
We present new analytical solutions for the evolution of protoplanetary discs (PPDs) where magnetohydrodynamic (MHD) wind-driven processes dominate.  Our study uses a 1D model which incorporates equations detailing angular momentum extraction by MHD winds and mass-loss rates. Our solutions demonstrate that the disc retains its initial state during the early phases; however, it rapidly evolves towards a self-similar state in the later stages of disc evolution. The total disc mass undergoes a continuous decline over time, with a particularly rapid reduction occurring beyond a certain critical time threshold. This gradual decrease in mass is influenced by the wind parameters and the initial surface density of the disc. In the MHD wind-dominated regime, we show that the disc's lifespan correlates positively with the magnetic lever arm up to a certain threshold, irrespective of the initial disc size. PPDs with a larger magnetic lever arm are found to maintain significantly higher total disc mass over extended periods compared to their counterparts. The mass ejection-to-accretion ratio increases in efficient wind scenarios and is further amplified by a steeper initial surface density profile. Our analysis also reveals varied evolutionary trajectories in the plane of accretion rate and total disc mass, influenced by magnetic parameters and initial disc size. In scenarios with efficient MHD winds, discs with bigger sizes have extended operation time for mechanisms governing planet formation. 

\end{abstract}


%%%%%%%%%%%%%%%%%%%%%%%%%%%%%%%%%%%%%%%%%%%%%%%%%%

%%%%%%%%%%%%%%%%% BODY OF PAPER %%%%%%%%%%%%%%%%%%

\section{Introduction}

Protoplanetary discs (PPDs), often regarded as nurseries where planets may form, have been at the forefront of astrophysical research in recent decades \citep[e.g.,][]{Andrews2020,Armitage2011}. As our observational capabilities have progressed, we have found details within these discs, enabling the development of increasingly sophisticated theoretical models that can be rigorously tested against observational evidence. These studies have greatly advanced our understanding of the physical conditions and processes within the birthplaces of planetary systems. One of the most intriguing and complex enigmas within this domain revolves around unravelling the dominant mechanisms governing accretion in PPDs. Over the years, various mechanisms have been suggested to explain accretion disc dynamics, including gravitational instability \citep[e.g.,][]{Lodato2004,Rafikov2009}, magnetorotational instability \citep{Balbus1991}, and hydrodynamic convection \citep[e.g.,][]{Kley1993, Held2018}. Yet, pinpointing the dominant mechanism through observational evidence continues to be a formidable task \citep[e.g.,][]{Simon2015, Rafikov2017, Wang2017,  Najita2018, Trapman2023b, Alexander2023, Wu2023}.

On the other hand, a substantial body of observational data concerning PPDs has emphasized the significant impact of magnetized and photoevaporative winds, ultimately resulting in the dispersal of these discs \cite[e.g.,][]{Pascucci2023}. Notably, theoretical investigations have illuminated the connection between turbulence generation, often driven by magnetorotational instability (MRI), and the concurrent production of magnetized winds \citep{Bai2013}. These winds not only induce mass loss but also significantly contribute to angular momentum removal. For many decades, the dominant perspective emphasized the primary role of turbulence-generating mechanisms, such as MRI, in driving the accretion process within disks. However, recent observational data have shifted this narrative,  indicating that MHD winds are pivotal in driving accretion within PPDs \citep[e.g.,][]{Weder2023}.

The launch of MHD winds from accretion discs has long been a subject of interest in astrophysical research. It was initially explored by \cite{Blandford1982}, who investigated the conditions required for the emergence of such winds using MHD self-similar solutions. Subsequent research expanded on this model through two primary approaches: on one hand, they explored MHD winds using analytical models \citep[e.g.,][]{Ferreira1993, Ferreira1997}, and on the other, they utilized sophisticated numerical simulations \citep{Bai2013}. These simulations, both with and without the inclusion of non-ideal factors such as Ohmic diffusion and the Hall effect, provided deeper insights into the complexities of MHD wind ejection.

The launch of MHD winds significantly impacts the structure of the underlying disc. However, exploring this interaction through global numerical simulations poses considerable challenges. Consequently, there is a compelling motivation to refine the classical accretion paradigm, as initially formulated by \cite{Shakura1973}, to study the pivotal role of MHD winds in determining the structure and dynamics of accretion discs.  These models typically incorporate equations describing angular momentum extraction by MHD winds and mass-loss rates, drawing inspiration from physical principles or the findings of numerical simulations. This approach not only facilitates a relatively straightforward analysis for investigating the structure and evolution of discs influenced by MHD winds but also offers results that can be compared against existing observational evidence of PPDs. Quantities such as disc size evolution, inner mass accretion rates, and total disc mass provide valuable insights that may yield criteria for distinguishing the dominant accretion mechanisms at play within PPDs.

Another significant research domain within the context of PPDs involves investigating the dynamics of dust particles and the mechanisms governing their growth. Understanding these processes is crucial for comprehending the intricate path to planet formation and the subsequent migration of planets within the disc. Although these complex issues have been subjects of study for many years, the potential impact of MHD winds has not been widely incorporated into previous research efforts. However, with recent advancements shedding light on the substantial role of MHD winds, there is an escalating interest in exploring these phenomena. Researchers are now employing both generalized $\alpha -$viscosity models \citep{Takahashi2018, Taki2021, Arakawa2021, Zagaria2022} and direct numerical simulations \citep[e.g.,][]{Kimmig2020, Aoyama2023} to probe into these problems and unveil their underlying mechanisms.

The ease of constructing generalized $\alpha -$viscosity models to incorporate MHD winds presents an advantage for exploring disc evolution across a wide range of input parameters. These models offer valuable physical insights into the expected behaviour of PPDs influenced by MHD winds. While the study by \cite{Suzuki2010} introduced a 1D disc model with MHD winds within the framework of the standard approach, it should be noted that their implementation of wind mass loss was based on MHD disc simulations. Building upon this work, \cite{Suzuki2016} further extended the model to incorporate both angular momentum removal and wind mass-loss rates. A noteworthy characteristic of their evolutionary models is that disc dispersal, driven by MHD winds, begins in the inner regions and progressively extends outward as time advances. 

A similar approach was undertaken by \cite{Armitage2013} to incorporate MHD winds into their model. They used relations for disc turbulence and angular momentum removal, based on plasma parameters derived from MHD simulations. This model was later extended by \cite{Shadmehri2019} to include wind mass-loss rates and an in-depth study of disc evolution was conducted. Various aspects, such as disc mass, net mass accretion rate at the inner disc edge, total wind mass-loss rate, and isochrone tracks \citep{Lodato2017}, were numerically calculated for different wind strengths. The isochrone tracks derived from their disc model, incorporating MHD winds, displayed reasonable fits with observations PPDs in star-forming regions like Lupus and $\sigma -$Orion.

A steady-state adaptation of this model was implemented by \cite{Khajenabi2018}, leading to analytical solutions that describe the consistent structure of PPDs with MHD winds. Their solutions also revealed an inner region characterized by a flat density profile, a consequence of the presence of MHD winds, in alignment with previous studies \citep[e.g.,][]{Suzuki2016}. These solutions have been utilized to explain the observational features of HL Tau, particularly its high accretion rate and the thinness of its dust layer \citep{Hasegawa2017}.

In parallel, several researchers have extended these disc models with MHD winds to include the dynamics of dust particles. For example, \cite{Takahashi2018} investigated the interplay between gas and dust in young PPDs, exploring how MHD wind launch processes contribute to the formation of ring-hole structures and the emergence of pressure bumps that migrate outward over time. Meanwhile, \cite{Taki2021} found a novel growth mode leading to the coalescence of dust grains into kilometre-sized bodies. This growth phenomenon, influenced by variations in the gas pressure profile induced by the wind, gives rise to ring-like configurations of planetesimal-sized entities within the inner regions of the disc, potentially shaping the subsequent planet formation process. Additionally, the outward radial drift of pebbles within PPDs has been attributed to the effects of MHD wind launch \citep{Arakawa2021}. Remarkably, in the context of circumplanetary discs, the emergence of MHD winds has also been shown to enhance the growth rate of dust particles through mutual collisions \citep{Shibaike2023}.

The majority of theoretical models for MHD wind-driven accretion discs, based on the generalized accretion disc model, are numerical in nature, lacking a closed analytical framework to describe disc evolution in the presence of MHD winds. However, a recent study by \cite{Tabone2022} (hereafter referred to as TB22) introduced an elegant set of similarity solutions for discs featuring MHD winds, characterized by a generalized form analogous to the well-known self-similar solutions for viscous discs originally proposed by \cite{Lynden-Bell1974}. These solutions have subsequently been applied to investigate the dynamics of dust particles within the context of MHD winds \citep{Zagaria2022} and to determine the theoretical size of MHD wind-driven accretion discs for comparison with observational data \citep{Trapman2022} and evolution of the distribution in the disk mass and accretion rate plane of a disk population \citep{Somigliana2023}. In addition, \cite{Chambers2019} also provided analytical solutions for modelling the evolution of discs influenced by MHD winds. While these solutions rely on simplifying approximations, they prove invaluable for conducting population synthesis studies related to planet formation \citep{Alessi2022}.

In this paper, we introduce {\it new} analytical solutions for the evolution of an accretion disc where MHD wind serves as the primary mechanism driving accretion. In this extreme configuration, inspired by observations of PPDs, as we will demonstrate, our model is integrable, allowing us to explore disc evolution as an initial value problem with a prescribed initial surface density distribution. While TB22 studied a wind-dominated disc model, their analysis was rooted in self-similar solutions, which inherently do not facilitate the tracking of disc evolution from a specified surface density distribution without further confirmation through numerical investigations. In the following section, we outline the fundamental equations of our model. Section 3 provides analytical solutions for MHD wind-driven accretion discs, using arbitrary initial surface density profiles, and extensively explores a comprehensive study of disc properties. Our findings are subsequently discussed in section 4.

\section{General Formulation}
Our model is based on the prior studies that investigated disc evolution with MHD winds, employing parameterized relationships for both angular momentum removal by MHD wind and wind mass-loss rate \citep{Suzuki2016, Armitage2013}. TB22 introduced slightly different formulations for these relations. In this work, we adopt their parameterizations to facilitate direct comparisons. This 1D model closely resembles the model for viscous evolution introduced by \citep{Shakura1973} and \citep{Lynden-Bell1974}. The primary evolutionary equation is as follows:
\begin{align}\label{eq:master}
\begin{split}
\frac{\partial\Sigma}{\partial t} = & \frac{3}{r}\frac{\partial}{\partial r} \left [ \frac{1}{r\Omega} \frac{\partial }{\partial r} \left (r^2 \alpha_{\rm SS} \Sigma c_{s}^2 \right ) \right ] 
 + \frac{3}{2r} \frac{\partial}{\partial r} \left ( \frac{\alpha_{\rm DW} \Sigma c_{s}^2}{\Omega} \right ) \\
& - \frac{3 \alpha_{\rm DW} \Sigma c_{s}^2}{4 (\lambda - 1 ) r^2 \Omega}.
\end{split}
\end{align}
In the above equation, \( \Sigma \) represents the disc surface density, while \( c_{s} \) denotes the sound speed. The Keplerian angular velocity is given by \( \Omega = \sqrt{GM/r^3} \), where \( M \) is the mass of the central star. The model incorporates three critical input parameters: \( \alpha_{\rm SS} \), which pertains to disc turbulence as introduced by \citet{Shakura1973}; \( \alpha_{\rm DW} \), associated with angular momentum removal by the wind, as discussed in TB22; and the magnetic lever arm parameter \( \lambda \), as described by  \citet{Blandford1982} and \cite{Ferreira1997}.

Our primary equation (\ref{eq:master}) involves two unknowns: surface density and sound speed. To close the system of equations, an energy equation is necessary. However, we pursue a simplified approach by adopting a power-law function of the radial distance for the disc temperature, or equivalently, sound speed. In this context, we assume that the sound speed is $c_{s}^{2} \propto r^{-1}$.

The mass accretion rate resulting from disc turbulence is expressed as
\begin{equation}\label{eq:Mvis}
\dot{M}_{\rm acc}^{\rm visc} = \frac{6\pi}{r\Omega} \frac{\partial}{\partial r} \left (\Sigma c_{s}^2 \alpha_{\rm SS} r^2 \right ).
\end{equation}
The mass accretion rate driven by MHD winds is formulated as
\begin{equation}\label{eq:Mdw}
    \dot{M}_{\rm acc}^{\rm DW}=\frac{3\pi \Sigma c_{s}^2 \alpha_{\rm DW}}{\Omega}
\end{equation}
The mass-loss rate due to MHD wind is parameterized as 
\begin{equation}\label{eq:sigmaW}
    \dot{\Sigma}_{\rm W} = \frac{3\alpha_{\rm DW} c_{s}^2}{4 (\lambda -1 ) \Omega r^2} \Sigma
\end{equation}

In the absence of MHD wind, the master equation (\ref{eq:master}) simplifies to the standard viscous evolutionary model \citep{Shakura1973, Lynden-Bell1974}. The second term on the right-hand side accounts for angular momentum removal by the MHD wind, which drives additional mass accretion beyond that induced by disc turbulence. The third term parameterizes the wind mass-loss rate, with larger \( \lambda \) values indicating reduced mass removal, essential for sustaining accretion onto the central star. Theoretical studies suggest a constraint of \( \lambda > 3/2 \) \citep{Blandford1982, Ferreira1997}.

\section{Analytical solutions}
As previously mentioned, our objective is to investigate the evolution of a disc when the contribution from disc turbulence to mass accretion is negligible, especially when compared to the influence of the MHD wind. Within the context of our model, this assumption translates to setting $\alpha_{\rm SS} = 0$. By introducing dimensionless variables, such as $\hat{\Sigma} = \Sigma / \Sigma_c$, $\hat{t} = t/t_c$, and $\hat{r} = r/r_c$, we can rewrite the master equation (\ref{eq:master}) as follows:
\begin{equation}\label{eq:master-wind}
\frac{\partial\hat{\Sigma}}{\partial\hat{t}}=\frac{1}{2\hat{r}}\frac{\partial}{\partial\hat{r}}\left ( \hat{r} \hat{\Sigma}\right ) - \frac{1}{4(\lambda -1)} \frac{\hat{\Sigma}}{\hat{r}}.
\end{equation}
We highlight that \(\Sigma_c\) and \(r_c\) represent the characteristic surface density and radial distance, respectively. For the characteristic time-scale \(t_c\), consistent with TB22, we adopt the accretion time-scale defined as 
\begin{equation}
  t_c \equiv t_{\rm acc, 0} =\frac{r_{c}^2}{3c_{s,c} H_{c} \alpha_{\rm DW}}.  
\end{equation}
In this expression, \(c_{s,c}\) and \(H_{c}\) denote the sound speed and disc thickness associated with the characteristic radial distance \(r_c\), respectively. We observe that by neglecting disc turbulence, the primary equation (\ref{eq:master}) undergoes a transformation. Specifically, it changes from a second-order partial differential equation to a first-order one, thereby simplifying its analysis. 

To solve equation (\ref{eq:master-wind}), we adopt a change of variables strategy. Upon applying the subsequent change of variables,
\begin{equation}
    \Sigma (\hat{r}, \hat{t}) = F(\hat{r}, \hat{t} ) \hat{r}^{-\frac{2\lambda -3}{2 \left (\lambda -1 \right )}}
\end{equation}
equation (\ref{eq:master-wind}) simplifies, facilitating a more straightforward analysis. We then obtain
\begin{equation}\label{eq:master-simply}
    \frac{\partial F}{\partial \hat{t}} - \frac{1}{2} \frac{\partial F}{\partial \hat{r}} =0.
\end{equation}
This is a first-order linear partial differential equation, commonly known as the transport or advection equation in one spatial dimension. It describes the evolution of a particular quantity, denoted by \(F(\hat{r}, \hat{t})\), which moves with a constant velocity of \(1/2\) in the radial direction. By employing the method of characteristics, this equation can be analytically solved, yielding the general solution:
\begin{equation}
    F(\hat{r}, \hat{t}) = F(\hat{r} + \frac{\hat{t}}{2} ),
\end{equation}
where \(F(\hat{r} + \frac{\hat{t}}{2} )\) can be any arbitrary function. However, it can be determined uniquely once the initial distribution of \(F\) is specified. Thus, the disc surface density becomes
\begin{equation}\label{eq:sol-f}
    \Sigma (\hat{r}, \hat{t}) = F(\hat{r}+\frac{\hat{t}}{2} ) \hat{r}^{-\frac{2\lambda -3}{2 \left (\lambda -1 \right )}}.
\end{equation}

We highlight that the fully analytical solution in equation (\ref{eq:sol-f}) is derived without imposing specific constraints on the density profile, as is typically done in similarity methods. Once the initial surface density distribution is established, we can use solution (\ref{eq:sol-f}) to determine \( F \) and, subsequently, deduce the disc evolution. 

As an example, consider employing an initial surface density given by
\begin{equation}\label{eq:initial}
    \Sigma (r , t=0) = \Sigma_c (\frac{r}{r_c})^{-n} e^{-m\frac{r}{r_c}},
\end{equation}
where $n$ and $m$ are free parameters. This particular choice is inspired by the self-similar viscous evolution of a disc presented by \cite{Lynden-Bell1974}. In our notation, their solution corresponds to $n=m=1$ with $\Sigma_c$ and $r_c$ as functions of time. TB22 introduced a profile analogous to that of equation~(\ref{eq:initial}) for disc evolution, subsequently aiming to derive the time dependencies \( \Sigma_c (t) \) and \( r_c (t) \). In the scenario of extreme wind dominance, their similarity solution indicated that \( n = (2\lambda -3)/[2(\lambda -1)] \). It is crucial to recognize that the self-similar approach is not inherently an initial value problem and, as such, does not incorporate the initial state adequately. Consequently, the initial surface density distribution in TB22 is contingent on the wind parameter \( \lambda \).

However, the potential roles of the MHD wind are examined once the initial configuration is established, independent of the wind parameters $\alpha_{\rm DW}$ and $\lambda$. Subsequently, we adjust these parameters to probe their influence on disc evolution. It is imperative to note that the initial state of the disc should remain uninfluenced by these wind parameters, ensuring that any alterations to them leave the initial configuration unchanged. Consequently, equation~(\ref{eq:initial}) presents the proposed initial surface density that aligns with our theoretical expectations and is suitable for our intended purpose. Using equations (\ref{eq:sol-f}) and (\ref{eq:initial}), we obtain
\begin{equation}\label{eq:surface-time}
    \Sigma (r,t) = \Sigma_{c} \left (\frac{r}{r_c} + \frac{t}{2t_{\rm acc,0}} \right )^{\frac{2\lambda -3}{2(\lambda -1)}-n} \left ( \frac{r}{r_c}\right )^{-\frac{2\lambda -3}{2(\lambda -1)}} e^{-m\left (\frac{r}{r_c} + \frac{t}{2t_{\rm acc,0}} \right )}.
\end{equation}
This solution characterizes the evolution of a PPD in the wind-dominated paradigm corresponding to any given exponent \(n\). Specifically, for \(n=(2\lambda -3)/[2(\lambda -1)] \), our solution becomes the self-similar solution deduced by TB22 (refer to eq.~(27) in TB22). This distinction implies that in the TB22 model, the initial configuration depends on the magnetic lever arm \(\lambda\), while in our approach, the exponent \(n\) for the initial state is a predetermined input. In our model, the primary input parameters encompass \( \Sigma_c \), \( r_c \), \( \lambda \), \( t_{\mathrm{acc, 0}} \), \( n \), and \( m \). The characteristic radius \( r_c \) typically ranges between 20 to 50 astronomical units (au). 

The surface density \( \Sigma_c \) can be deduced from the initial disc mass by utilizing the surface density profile as presented in equation \eqref{eq:initial}. This relationship can be written as
\begin{equation}\label{eq:integral}
    \Sigma_c = \frac{M_{\mathrm{D0}}}{2\pi r_{c}^2 I(\hat{r}_{\mathrm{in}}, n, m)},
\end{equation}
where, \( I(\hat{r}_{\mathrm{in}},n,m) \) is given by 
\begin{equation}
  I(\hat{r}_{\mathrm{in}},n,m) = \int_{\hat{r}_{\mathrm{in}}}^{\hat{r}_{\rm out}} \hat{r}^{1-n} \exp(-m \hat{r}) \, d\hat{r}  
\end{equation}
 and \( \hat{r}_{\mathrm{in}} = r_{\mathrm{in}}/r_c \) and \( \hat{r}_{\mathrm{out}} = r_{\mathrm{out}}/r_c \). Notably, \( {r}_{\mathrm{in}} \) represents the inner disc radius, and it's typically much smaller than \( r_c \). As for the outer edge of the disc, we generally assume a large value, say, $r_{\rm out} = 1000$ au. Furthermore, the term \( M_{\mathrm{D0}} \) represents the initial disc mass. For typical protoplanetary discs (PPDs), this mass usually falls within the range \( 10^{-3} \) to \( 10^{-2} \) M\(_\odot\).

\begin{figure}
\includegraphics[scale=0.55]{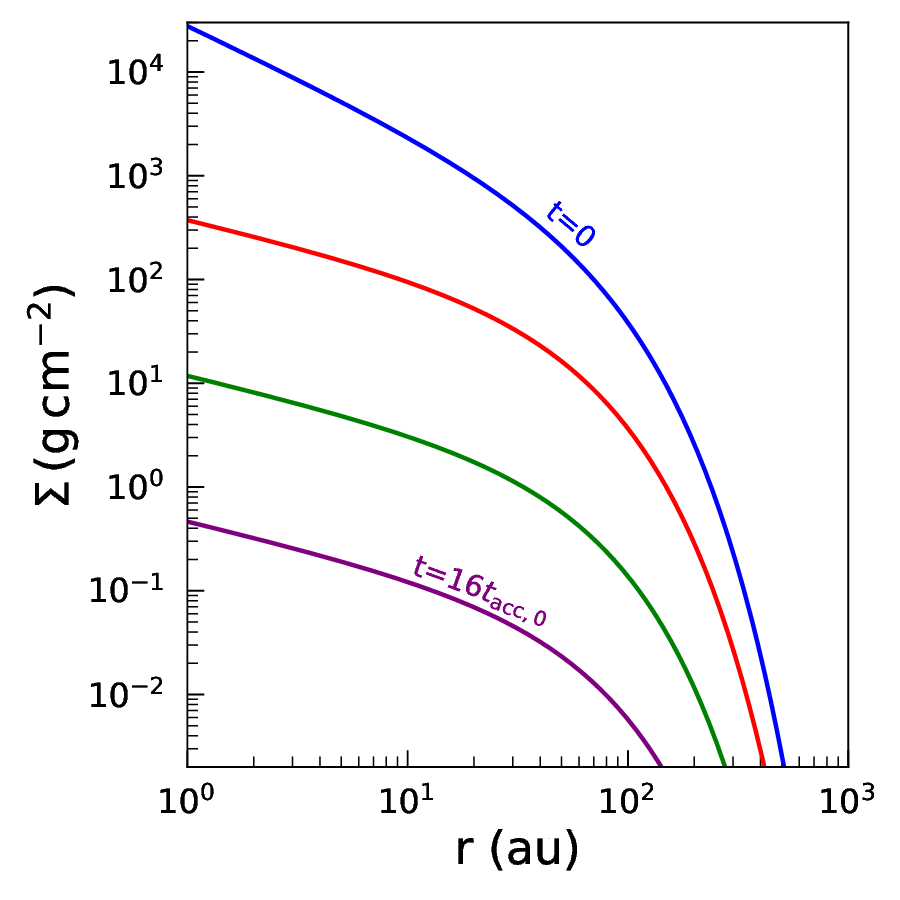}
\caption{Evolution of the surface density as a function of radial distance based on our analytical solution for the wind-dominated disc with input parameters \( r_c = 50 \) au, \( M_{D0} = 0.01 \) M\(_\odot\), \( m = n = 1 \), and \( \lambda = 2 \). The curves, from top to bottom, correspond to time instances: \( t = 0 \), \( 4 t_{\mathrm{acc,0}} \), \( 10 t_{\mathrm{acc,0}} \), and \( 16 t_{\mathrm{acc,0}} \), respectively.}
\label{fig:f1}
\end{figure}

Utilizing the solution presented in equation (\ref{eq:surface-time}), we initiate a comprehensive investigation into the characteristic evolutionary dynamics embedded within our disc model. Figure \ref{fig:f1} offers a view of the surface density's evolution, depicting its progression over a range of times, which are expressed in units of \(t_{\rm acc,0}\). The initial disc mass is set at \(10^{-2} M_\odot\) and the characteristic radius \(r_c\) is chosen to be 50 au. In addition, we set the remaining input parameters to \(n=m=1\) and \(\lambda=2\). In Figure \ref{fig:f1}, the evolutionary curves for \( t = 0 \), \( 4 t_{\mathrm{acc,0}} \), \( 10 t_{\mathrm{acc,0}} \), and \( 16 t_{\mathrm{acc,0}} \) are displayed in descending order from top to bottom.

A key feature in Figure \ref{fig:f1} is the consistent decrease of the surface density across all radii as time progresses. Over a span of $16 t_{\rm acc,0}$, the reduction in surface density is significant, approximating a factor of 10,000. The reduction of the disc surface density is attributed to the dual effects of wind-driven accretion onto the star and the wind-induced mass loss. Initially, the density distribution manifests as particularly steep, especially in the disc's outer regions. However, as time elapses, the surface density profile adopts a more moderate incline. This moderation initiates from the innermost regions and systematically extends outwards. This inside-out manner of gas removal remains largely unaltered across varying model parameters. While the early evolution of the disc is influenced by the initially chosen configuration, over time, the disc tends toward a self-similar profile in its later stages. The characteristic time-scale $t_{\text{acc},0}$ provides a measure of the disc's time-scale evolution. For periods longer than $t_{\text{acc},0}$, the disc evolution exhibits self-similar behavior.

\begin{figure}
\includegraphics[scale=0.55]{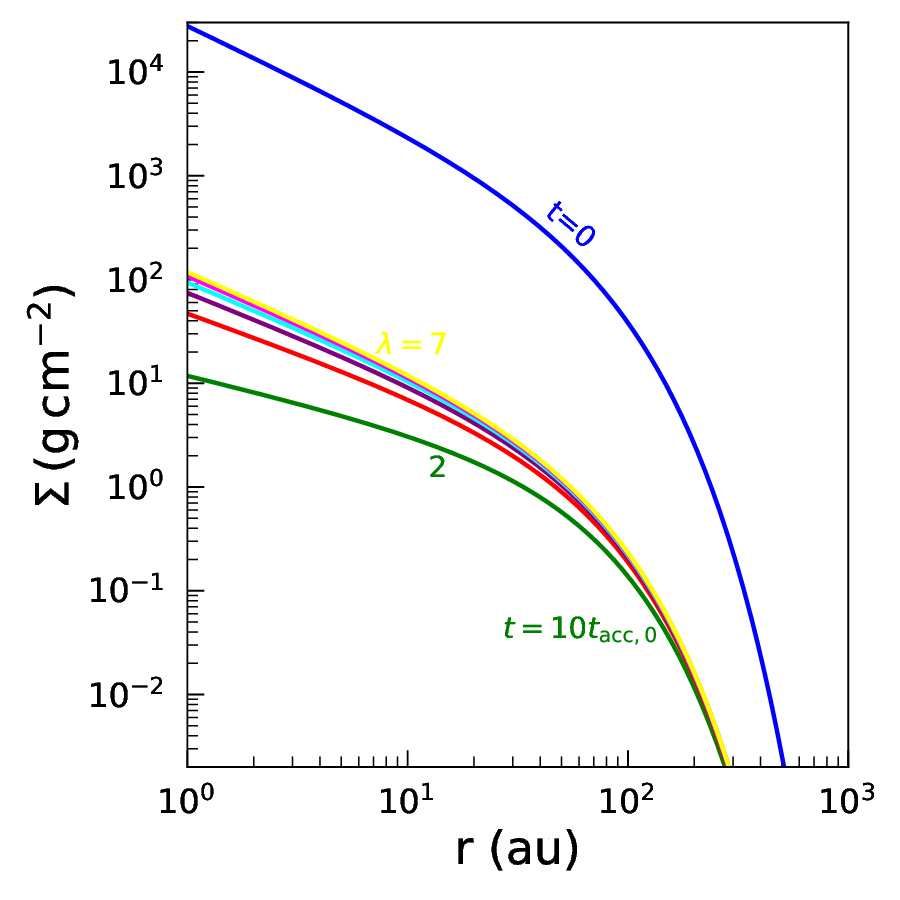}
\caption{This figure illustrates the initial and subsequent surface density profiles across radial distances, emphasizing the role of the wind parameter \( \lambda \). All the input parameters remain consistent with those specified in Figure \ref{fig:f1}. While the profile at \( t = 0 \) remains unchanged, for \( t = 10 t_{\mathrm{acc,0}} \), the surface density profiles are  presented for different \( \lambda \) values, ranging from \( \lambda = 2 \) at the bottom to \( \lambda = 7 \) at the top. }
\label{fig:f2}
\end{figure}

In the context of MHD wind mass-loss, the magnetic lever arm parameter \(\lambda\) serves as a measure of efficiency. An increase in \(\lambda\) results in a reduced rate of wind mass-loss to maintain the needed mass accretion rate onto the star. Figure~\ref{fig:f2} displays the influence of the wind parameter \(\lambda\) on the evolution of the surface density. The considered input parameters align with those in figure~\ref{fig:f1}. However, in this depiction, the surface density profile is consistently captured at \(10 t_{\mathrm{acc, 0}}\), spanning varying \(\lambda\) values---ascending from \(2\) at the bottom to \(7\) at the top. A clear pattern is evident: as \(\lambda\) increases, the decline in surface density becomes less prominent, consistent with our anticipations. While the outer regions of the disc largely resist alterations from this parameter, the inner domains display marked sensitivity. However, it's worth noting that as \(\lambda\) exceeds an approximate value of 5, its subsequent impact on the surface density becomes considerably reduced.

We have also examined scenarios with varying values of \(n\) and \(m\), which correspond to different initial surface density profiles. The observed trends in these cases align closely with the results presented thus far. While we have chosen not to detail these specific cases here, the influence of the exponent $n$ on disc quantities, including total disc mass and the mass accretion rate, will be discussed in the subsequent figures.

Given that our model incorporates MHD wind as the dominant mechanism driving both mass accretion onto the host star and mass loss, it is expected that the total disc mass will decrease over time. To quantify this theoretical expectation, we employ the surface density profile outlined in equation (\ref{eq:surface-time}) to compute the total disc mass. The total disc mass is derived as follows
\begin{equation}\label{eq:Mdt}
    M_{\rm D}(t) = \frac{M_{\rm D0}}{I(\hat{r}_{\rm in}, n, m)} \int_{\hat{r}_{\rm in}}^{\hat{r}_{\rm out}} f(\hat{r}, \hat{t}) \hat{r} d\hat{r},
\end{equation}
where $I(\hat{r}_{\rm in}, n, m)$ is given by equation (\ref{eq:integral}) and the function $f(\hat{r}, \hat{t}) $ is
\begin{equation}
  f(\hat{r}, \hat{t}) =  \left (\hat{r} + \frac{\hat{t}}{2} \right )^{\frac{2\lambda -3}{2(\lambda -1)}-n}  \hat{r}^{-\frac{2\lambda -3}{2(\lambda -1)}} e^{-m\left (\hat{r} + \frac{\hat{t}}{2} \right )}.
\end{equation}

\begin{figure}
\includegraphics[scale=0.55]{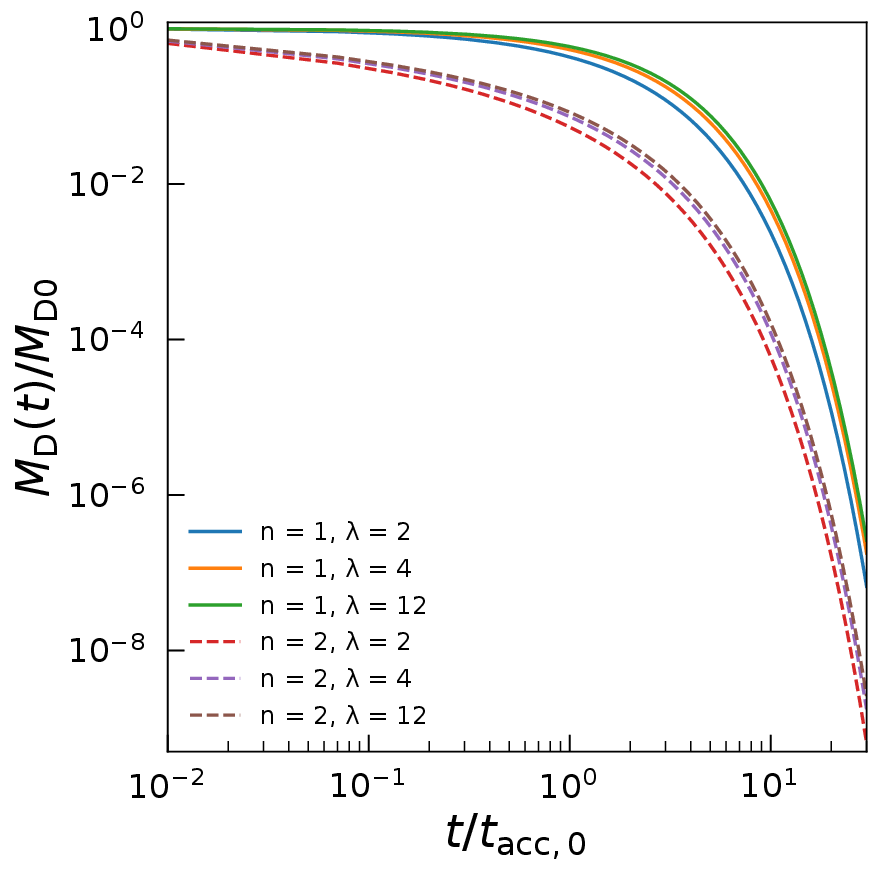}
\caption{Evolution of the normalized total disc mass over time for varying values of \( \lambda = 2, 4, 12 \) and exponents \( n = 1 \) and \( 2 \). The total disc mass is normalized by its initial value. The input parameters, unless otherwise mentioned, remain consistent with those presented in Figure \ref{fig:f1}. Regardless of the chosen input parameters, the total disc mass exhibits a decline over time.}
\label{fig:f3}
\end{figure}

Figure \ref{fig:f3} presents the evolution of the total disc mass normalized by its initial value over time. The ratio $M_{\rm D}(t) / M_{\rm D0}$ is shown for two distinct initial configurations, characterized by \( n = 1 \) and \( n = 2 \). For each value of \( n \), we also explore various values of \( \lambda \) as shown in the figure. The disc's inner edge is consistently set at \( r_{\rm in} = 0.05 \) au.

A general trend evident across all examined cases is the anticipated decline in total disc mass with time. While this decline is initially moderate, an approximate distinct point can be observed beyond which the reduction becomes sharply accelerated. This suggests a two-phase evolutionary sequence for the disc: the early phase is marked by a measured depletion of mass, whereas the subsequent phase witnesses a precipitous drop within a limited time span. Notably, similar trends have been highlighted in prior studies \citep[e.g.,][]{Suzuki2016, Shadmehri2019,  Zagaria2022}

A decrease in \( \lambda \)  further amplifies the reduction in disc mass. Yet, a particular point of intrigue emerges when considering \( n = 2 \), representing a considerably steeper initial configuration. In this case, the gradual phase is shortened significantly, leading to an earlier onset of rapid decline. Given that larger \( n \) values, assuming constant initial mass, suggest smaller disc sizes, it can be inferred that steeper initial surface density profiles accelerate the disc's depletion rate. Consequently, the lifespan of the disc is substantially reduced for steeper profiles. These observed trends diverge from the findings of TB22. While they also identified an exponential decrease in total mass, their results were not influenced by the magnetic lever arm or the slope of the initial surface density.

\begin{figure}
\includegraphics[scale=0.55]{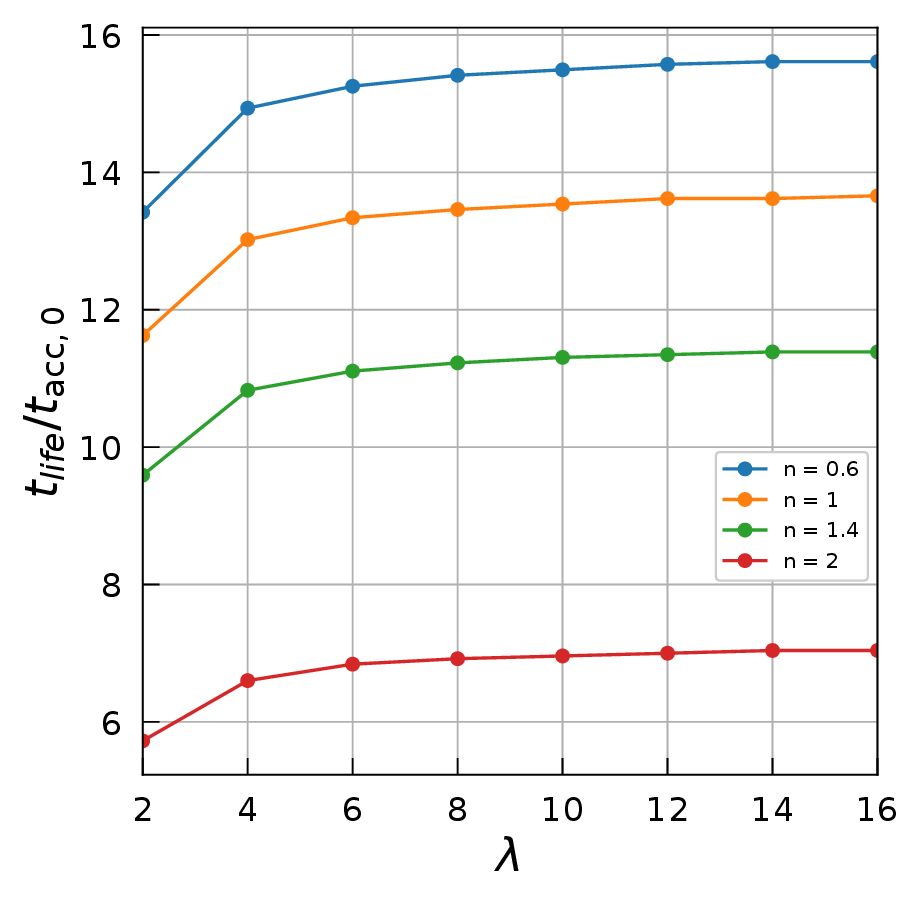}
\caption{In this figure, we present the normalized disc lifetime $t_{\rm life}$, scaled by \( t_{\mathrm{acc,0}} \), plotted against the wind parameter \( \lambda \) for various values of \( n \) as indicated. The disc lifetime is defined at the point where the disc mass falls below a specified threshold, in this instance, \( 10^{-3} \) of its initial mass.}
\label{fig:f4}
\end{figure}

In Figure \ref{fig:f4}, we explore the disc's lifetime $t_{\rm life}$ across a range of values for both \(\lambda\) and \(n\). In our analysis, we define the disc's lifetime as the moment when its mass diminishes to a predefined fraction of its initial mass. It is worth noting that while we adopt a threshold fraction of \(10^{-3}\) in this study, the general trends and behaviour of the disc's lifetime, in relation to the input parameters, remain largely consistent irrespective of the specific choice of this threshold. Figure \ref{fig:f4} shows that the disc lifetime increases as the magnetic lever arm increases, but beyond $\lambda \simeq 6$, this lifetime remains nearly uniform no matter what exponent $n$ is adopted.

\begin{figure}
\includegraphics[scale=0.55]{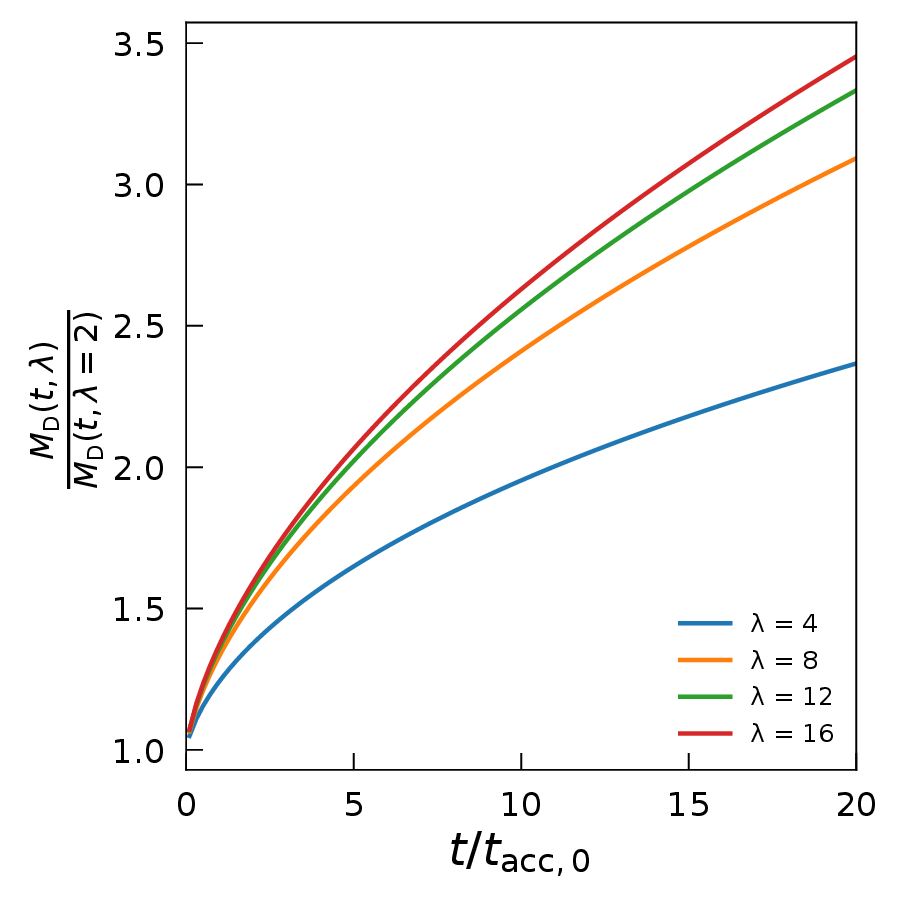}
\caption{Evolution of the mass ratio \( M_{\rm D} (t, \lambda)/M_{\rm D} (t, \lambda=2) \) for \( n=1 \). As \( \lambda \) increases, a larger fraction of the mass remains in the disc. Over time, this difference in retained mass becomes more pronounced.}
\label{fig:f5}
\end{figure}

\begin{figure}
\includegraphics[scale=0.55]{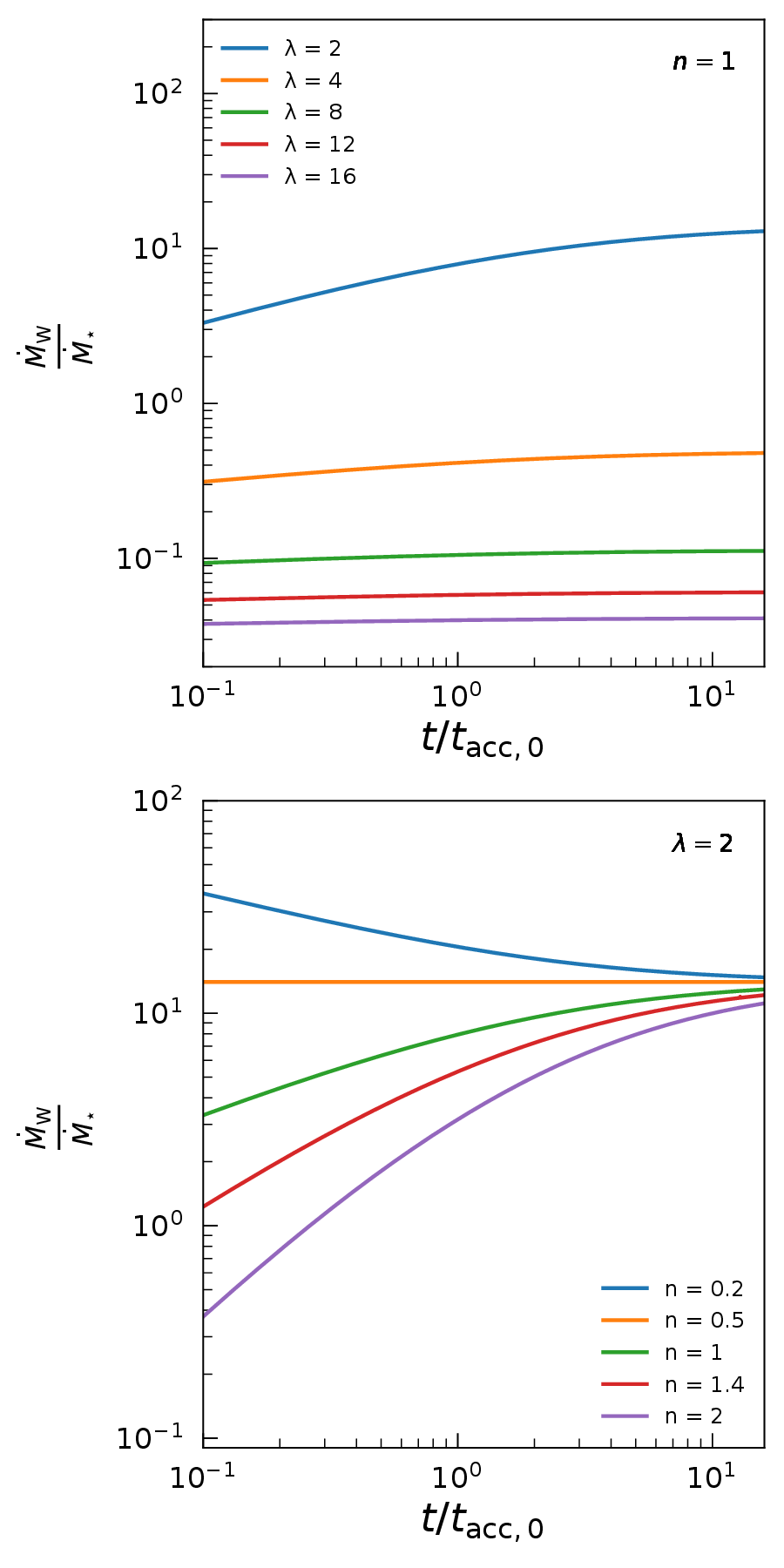}
\caption{Evolution of the mass ejection-to-accretion ratio with time is shown for different input parameters. The top plot displays the ratio for various magnetic lever arm parameter values, \( \lambda = 2, 4, 8, 12, 16 \), highlighting the influence of \( \lambda \) on mass depletion dynamics. The bottom plot contrasts this by focusing on a fixed \( \lambda = 2 \) and varying the exponent \( n = 0.2, 0.5, 1, 1.4, 2 \), emphasizing the role of the initial disc size in the ratio's temporal behaviour. The above trends deviate from the uniform behaviour predicted by the self-similar solution presented by TB22.
}
\label{fig:f6}
\end{figure}

Figure \ref{fig:f5} offers a comparative analysis of the total disc mass for various values of \( \lambda = 4, 8, 12, 16 \) relative to the baseline case of \( \lambda = 2 \). This comparison aims to underscore the pivotal role played by the magnetic lever arm parameter, \( \lambda \), in influencing the effectiveness of mass removal via wind-driven processes. Contrasting the evolution trajectories, it becomes evident that the time-dependent ratio \( M_{\rm D} (t, \lambda)/M_{\rm D} (t, \lambda=2) \) for \( n=1 \) increases with time. Furthermore,  discs associated with lower \( \lambda \) values experience a more pronounced wind-driven mass extraction compared to their counterparts with higher \( \lambda \) values. To put this into perspective, by the epoch of \( 20 t_{\rm acc,0} \), the disc mass corresponding to \( \lambda = 4 \) is approximately double that of the \( \lambda = 2 \) case. Simultaneously, the disc mass associated with \( \lambda = 8 \) is found to be three times that of the \( \lambda = 2 \) case. This comparison effectively illustrates the intricate interaction between \( \lambda \) and the dynamics of wind-driven mass depletion in PPDs.

In the framework of our model, mass reduction of the disc is primarily due to the accretion onto the central star and mass ejection through MHD winds. Now, we precisely quantify these mass loss rates. 
Using equations \eqref{eq:Mvis} and \eqref{eq:Mdw}, we can now determine the net mass accretion rate onto the star, denoted by $\dot{M}_{\star}$, which is expressed as
\begin{equation}
  \dot{M}_{\star} = \dot{M}_{\rm acc}^{\rm vis} (r_{\rm in}) + \dot{M}_{\rm acc}^{\rm DW} (r_{\rm in}).  
\end{equation}
It is important to note that both terms are evaluated at the inner disc edge. In the wind-dominated regime, however, the contribution of the first term is zero. Therefore, we obtain the following expression for $\dot{M}_{\star}$:
\begin{equation}\label{eq:Mstar}
    \dot{M}_{\star}(t) = \frac{M_{\rm D0}}{t_{\rm acc,0}} \frac{\hat{r}_{\rm in}}{2 I(\hat{r}_{\rm in}, n, m)} f(\hat{r}_{\rm in}, \hat{t}).
\end{equation}

Using the prescribed local mass loss rate, as given by equation \eqref{eq:sigmaW}, we can calculate the cumulative mass-loss rate across the entirety of the disc. This rate, denoted by \(\dot{M}_{\rm W}\), is formulated through the integral of the local rate over the radial extent of the disc, spanning from its inner edge, \(r_{\rm in}\), to its outer boundary, \(r_{\rm out}\). This cumulative rate is given by the relation 
\begin{equation}
\dot{M}_{\rm W} (t) = 2\pi \int_{r_{\rm in}}^{r_{\rm out}} \dot{\Sigma}_{\rm W} r dr.
\end{equation}
Thus, we obtain
\begin{equation}
\dot{M}_{\rm W} (t) = \frac{M_{\rm D0}}{t_{\rm acc, 0}} \frac{1}{4(\lambda -1)}\frac{1}{I(\hat{r}_{\rm in}, n, m)} \int_{\hat{r}_{\rm in}}^{\hat{r}_{\rm out}} f(\hat{r}, \hat{t}) \hat{r} d\hat{r}.
\end{equation}

A particularly suitable parameter for assessing the influence of MHD winds within the disc is the dimensionless mass ejection-to-accretion ratio, denoted as \(\dot{M}_{\rm W}/\dot{M}_{\star}\) \citep[e.g.,][]{Ferreira1997}. One of the notable advantages of this parameter is its observability, in addition to its independence from the disc mass. TB22 derived an expression for this ratio utilizing their self-similar solution. In a case where the wind dominates, they found this ratio as $(r_{\rm c}/r_{\rm in})^{1/[2(\lambda -1)]} -1$. But our solution implies that
\begin{equation}
    \frac{\dot{M}_{\rm W} (t)}{\dot{M}_{\star} (t)} = \frac{1}{2(\lambda -1) \hat{r}_{\rm in}} \frac{1}{ f(\hat{r}_{\rm in}, \hat{t})} \int_{\hat{r}_{\rm in}}^{\hat{r}_{\rm out}} f(\hat{r}, \hat{t}) \hat{r} d\hat{r}.
\end{equation}
The ratio is dependent upon the input parameters and, as we will demonstrate, can vary over time.

The top plot in Figure \ref{fig:f6} illustrates the mass ejection-to-accretion ratio $ \dot{M}_{\rm W} (t)/ \dot{M}_{\star} (t)$ as a function of time, considering various values of \( \lambda = 2, 4, 8, 12, 16 \) from the top curve to the bottom curve, respectively. It is evident that for \( \lambda \) values greater than 4, the ratio remains below unity throughout the entire evolution of the PPD. Conversely, for \( \lambda \) values less than 4, the ratio exceeds unity. In the later case, the ratio demonstrates a non-uniform behaviour over time, which contrasts the expectations based on the self-similar solution presented by TB22. The increase in the mass ejection-to-accretion ratio over time indicates that the rate of wind mass loss becomes greater than the mass accretion rate onto the central star.

The bottom plot of Figure \ref{fig:f6} exhibits the evolution of the mass ejection-to-accretion ratio for a fixed \( \lambda = 2 \), while considering different values of the exponent \( n = 0.2, 0.5, 1, 1.4, 2 \), from the top curve to the bottom curve, respectively. Remarkably, while the ratio remains relatively uniform over time for \( n = 0.5 \), for other \( n \) values, the ratio displays substantial temporal variation. This distinctive trend and evolution of the mass ejection-to-accretion ratio are not captured by the self-similar solution presented by TB22, where a uniform ratio was obtained for all input parameters. In contrast, Figure \ref{fig:f6} (bottom) clearly illustrates that the ratio increases with time when the exponent \( n \) exceeds  0.5. However, for \( n < 0.5 \), the trend is reversed, and the ratio decreases with time. Our model suggests that the initial size of the disc plays a pivotal role in determining this ratio's behaviour. In smaller discs, the ratio gradually increases over time, while in larger discs, the ratio declines.

\begin{figure}
\includegraphics[scale=0.55]{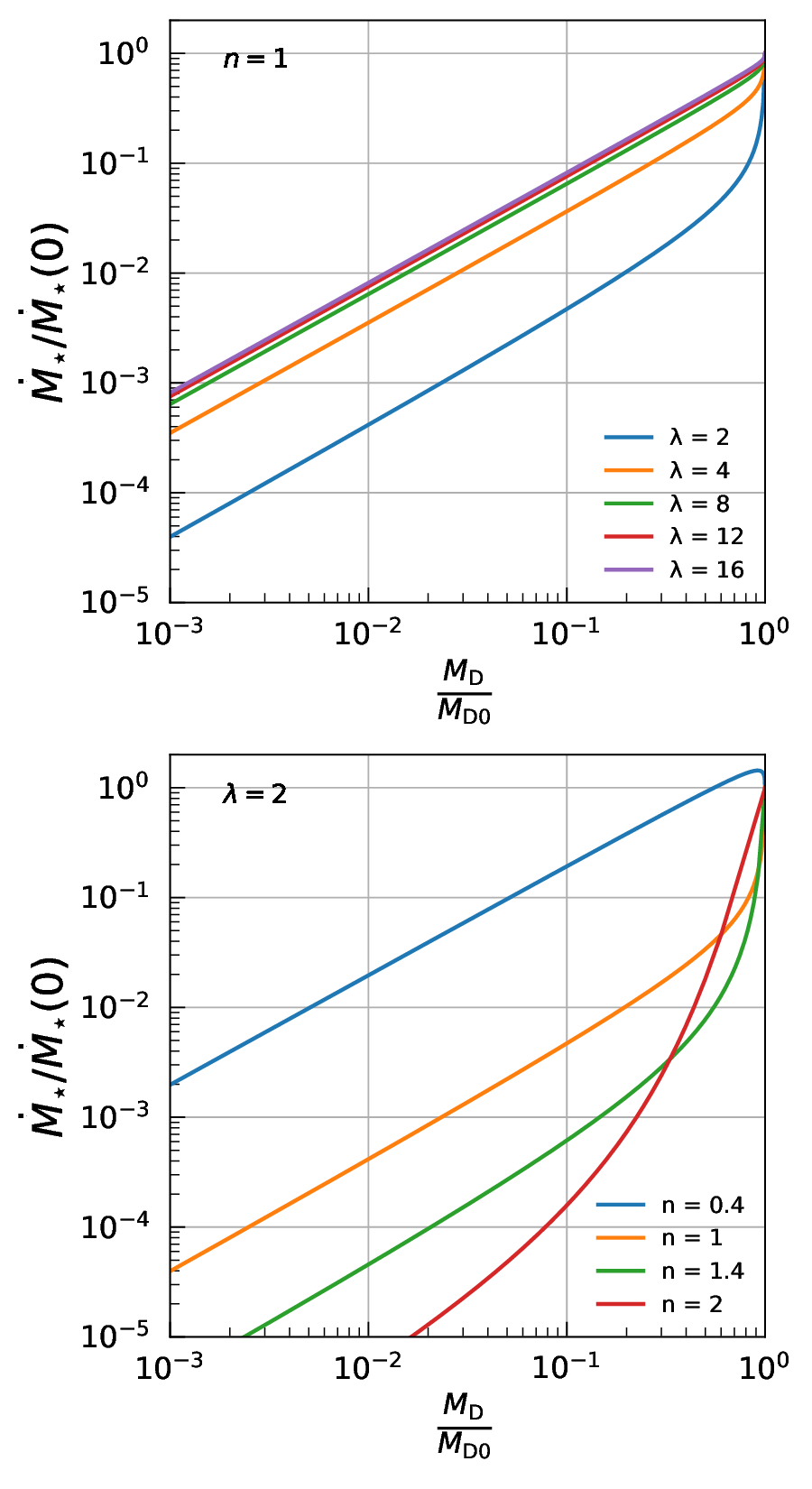}
\caption{Evolutionary trajectories of PPDs in the \(\dot{M}_{\star}-M_{\rm D}\) plane. The top plot displays trajectories for constant \( n = 1 \) with varying magnetic lever arm parameter \( \lambda \), showing diverse characteristics based on \( \lambda \) values. In contrast, the bottom plot, set at a constant \( \lambda = 2 \), explores the impact of differing exponent \( n \) values on these trajectories, highlighting the influence of the initial surface density distribution slope. As evolution unfolds, trajectory slopes from both plots exhibit a tendency for convergence.}
\label{fig:f7}
\end{figure}

In Figure \ref{fig:f7}, we explore the evolution of PPDs, not only focusing on changes in the total disc mass but also variations in the mass accretion rate at the inner disc edge. Similar to previous studies, to characterize this evolution, we introduce a plane with the vertical axis represented by \( \dot{M}_{\star}(t) / \dot{M}(t=0) \) and the horizontal axis by \( M_{\rm D}(t) / M_{\rm D}(t=0) \). Each snapshot in time corresponds to a distinct point on a plane $\dot{M}_{\star}-M_{\rm D}$, together forming a trajectory that chronicles the disc's progression. Although our surface density profile is analytical, the disc's trajectory in this plane is determined numerically using equations (\ref{eq:Mdt}) and (\ref{eq:Mstar}).

In the top plot of Figure \ref{fig:f7}, the trajectories of disc evolution are shown for a constant \( n = 1 \) across varying magnetic lever arm parameter \( \lambda \) values, specifically \( \lambda = 2, 4, 8, 12, 16 \), sequenced from bottom to top. Notably, TB22, in their analysis, employed their self-similar solutions to deduce these trajectories, uncovering a linear relation in wind-dominated case,  i.e. \( \dot{M}_{\star} \propto M_{\rm D} \), independent of the \( \lambda \) parameter. However, our findings in figure \ref{fig:f7} underscore that trajectories can exhibit diverse characteristics depending on the value of \( \lambda \). In scenarios where wind mass removal is markedly efficient (lower $\lambda$ values), the trajectory's slope, especially in the nascent stages of disc evolution, is steeper compared to discs with diminished mass removal rates (higher $\lambda$ values). Yet, as the timeline progresses, a convergence in slope is evident across the explored cases. For more substantial \( \lambda \) values, there is a discernible downward shift in the trajectory, suggesting a reduced inner accretion rate for equivalent total disc mass.

The bottom plot of figure \ref{fig:f7} examines cases with a constant \( \lambda = 2 \), exploring variations in the exponent \( n \) with values \( n = 0.4, 1, 1.4, 2 \).  Our analysis reveals that the trajectory within the \( \dot{M}_{\star}-M_{\rm D} \) plane, as the disc evolves temporally, is considerably influenced by the initial slope of the surface density distribution. As noted earlier, a rise in \( n \) for a specified total mass implies a reduced disc size. This is particularly evident when size is determined by a threshold surface density metric \citep[e.g.,][]{Toci2023} or by the radius encompassing a certain fraction of the disc's mass \citep[e.g.,][]{Anderson2013}. As a result, smaller initial disc sizes lead to steeper trajectories with a downward shift. Conversely, larger initial disc sizes produce an upward shift in the trajectory, aligning more closely with a power-law relationship. Interestingly, during the later stages of evolution, the trajectory slopes converge to the same value, independent of the adopted values for the exponent \( n \).

We have shown corresponding curves in Figure \ref{fig:f7} using dimensionless physical quantities, providing a convenient framework for a more straightforward interpretation of theoretical trends. However, to facilitate a meaningful comparison with observational data, it is necessary to illustrate these trends in the $\dot{M}_{\star} - M_{D}$ plane using actual physical units. It is essential to acknowledge that a comprehensive understanding of the conditions required to align these curves with observations necessitates a detailed exploration through realistic disc population synthesis, a task that lies beyond the scope of the present study.

\begin{figure}
\includegraphics[scale=0.55]{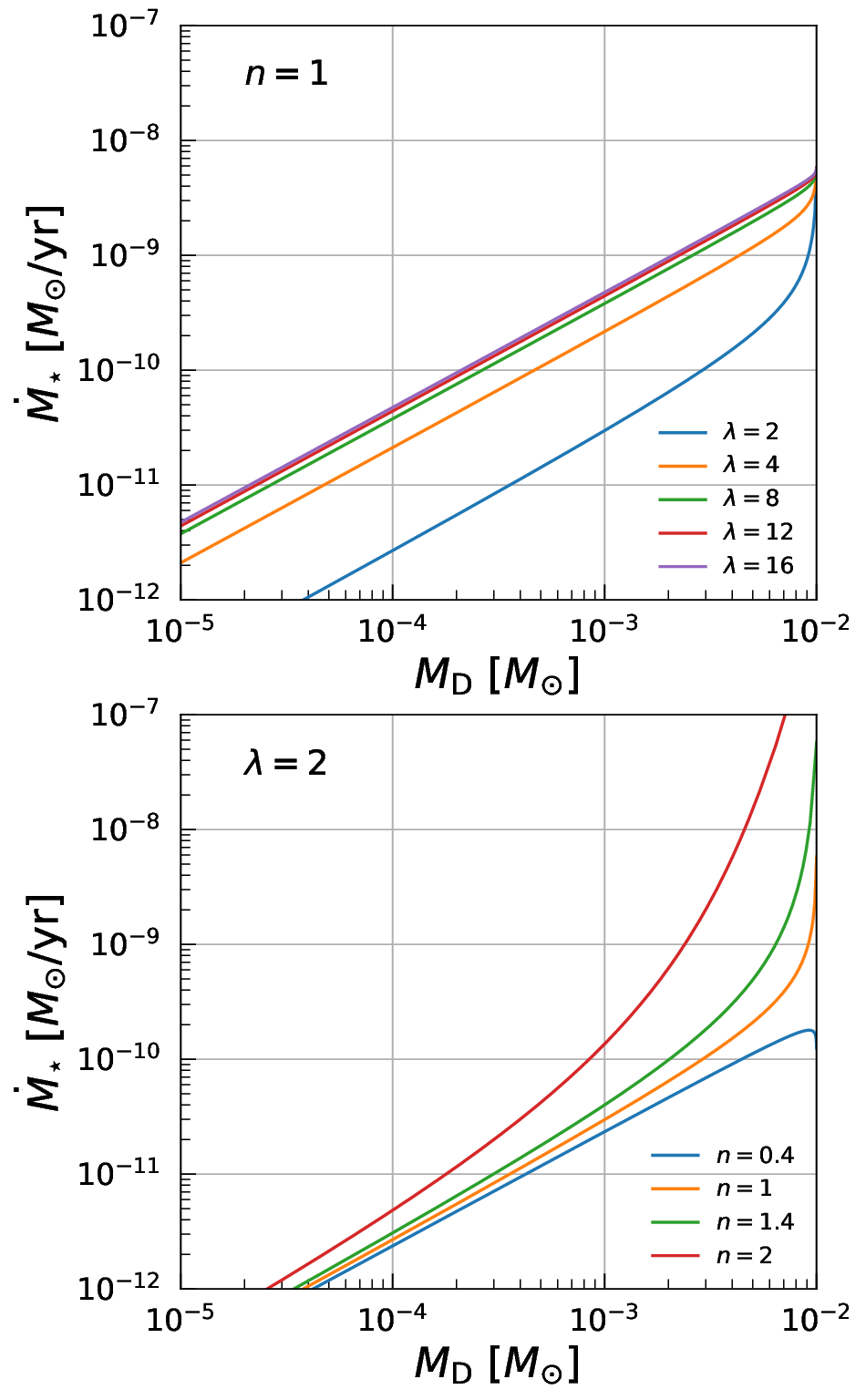}
\caption{This figure is identical to Figure \ref{fig:f7}, with the distinction that the presented curves represent the evolutionary trajectories in the $\dot{M}_{\star} - M_{D}$ plane using actual physical units. These curves correspond to an ensemble of discs characterized by an initial disc mass of $M_{D0}=0.01$ M$_{\odot}$ and an accretion timescale of $t_{\text{acc, 0}}=1$ Myr.}

\label{fig:f8}
\end{figure}

Figure \ref{fig:f8} is similar to Figure \ref{fig:f7}, with the distinction that we now present the curves in terms of actual physical units. The initial disc mass is $M_{D0} = 0.01$ M$_{\odot}$. As previously discussed in TB22, typical values for the accretion timescale, $t_{\text{acc, 0}}$, are of the order of a million years (Myr). In this context, we adopt $t_{\text{acc, 0}} = 1$ Myr.

In both Figure \ref{fig:f7} and \ref{fig:f8}, we can easily recognize an initial phase of disc evolution followed by a profile corresponding to self-similar evolution. The duration of the early disc phase is dependent on the wind parameter $\lambda$ and the initial surface density exponent $n$. As demonstrated in the total disc mass evolution presented in Figure \ref{fig:f3}, we have previously established that disc evolution commences with an early phase, followed by a more extended, long-term evolutionary phase. During the early disc phase evolution, the total disc mass undergoes only a modest reduction. However, beyond a certain time, the disc mass experiences a rapid decline.

Figure \ref{fig:f3} illustrates that the duration of the initial phase is of the order of the initial accretion timescale $t_{\text{acc, 0}}$. In this initial disc phase evolution, the total disc mass does not decrease rapidly, as depicted in both Figure \ref{fig:f7} and \ref{fig:f8} by the initial segments that deviate from the self-similar profile. In the case with $n=1$ and $\lambda=2$, the duration of the initial phase is slightly longer, though it remains shorter than the initial accretion timescale.

For a fixed $\lambda=2$ and different values of $n$, the longest duration of the initial phase belongs to $n=2$. Notably, during the early phase in this specific case, the total disc mass is reduced by a factor of $10^2$. Figure \ref{fig:f3} indicates that it takes a few initial timescales to achieve this level of mass reduction. Therefore, it appears that the initial evolutionary phase is significantly shorter than the entire lifetime of the disc.

\section{Discussion and Conclusions}

Our analytical solutions for the evolution of MHD wind-dominated discs demonstrate that the initial disc configuration plays a pivotal role in subsequent disc evolution. When disc turbulence is parameterized by a viscous term, the disc evolves towards a state independent of the initially adopted surface density profile. In this scenario, it's as if the disc forgets its initial state due to the efficient spreading caused by viscous processes. However, in the absence of disc turbulence, the nature of the master equation changes, as we have already discussed. In such a case, 
the disc retains its initially chosen configuration in $t<t_{acc,0}$, but subsequently the solution converges to the self-similar one. In our study, we have utilized a commonly adopted initial surface density profile, and our results differ from those presented by TB22. It's important to note that the exact evolutionary sequence of PPDs, including the onset of MHD wind launching, remains a subject of uncertainty.

During the early stages of PPD formation, these discs tend to be more massive and susceptible to gravitational instability as the dominant mechanism for angular momentum transfer \cite[e.g.,][]{Kratter2016}. It remains uncertain whether, as the disc depletes its mass, it undergoes a transition into an MRI-driven accretion phase, with or without the influence of MHD winds. For a specific choice of $n=1$ in our model, we have assumed that the disc starts its evolution driven by viscous mechanisms, ultimately adopting a self-similar profile similar to that described by \cite{Lynden-Bell1974} when the wind-dominated regime commences.  It is at this stage that MHD wind becomes efficient, justifying our choice of initial configuration with $n=1$. It is important to recognize that this scenario regarding the adopted initial state is speculative, and further extensive studies are warranted to explore the diverse phases of PPD evolution.

Dispersal of PPDs, whether due to MHD winds or photoevaporative winds, remains a critical area of research \cite[e.g.,][]{Kunitomo2020,Pascucci2023}. \cite{Kunitomo2020} investigated the impact of both MHD and photoevaporative winds by incorporating mass loss rates associated with the winds and considering angular momentum extraction by MHD winds, employing a generalized standard disc model. Their findings reveal that, during the early stages of disc evolution, MHD winds are efficient in the inner disc region. However, as the evolution progresses, photoevaporative winds become dominant, particularly in the outer disc region and during the later phases of disc evolution.  In our analytical model, we explored an extreme scenario where MHD winds are the dominant factor in disc dispersion. Within this framework, disc lifetime is quantified as the period during which the disc loses a substantial portion of its initial mass. Our findings suggest that the total disc mass decreases over time, with its evolution being intricately tied to the magnetic lever arm and the gradient of the initial density profile. This contrasts with the findings of TB22, who observed an exponential decline regardless of wind parameters and initial conditions based on their similarity solutions. In our wind-dominated solutions, smaller discs exhibit a reduced lifetime. It's important to emphasize that this characteristic trend emerges in the absence of disc turbulence, which typically promotes disc spreading.

As mentioned earlier, discriminating between the dominant accretion processes in PPDs, whether driven by viscous evolution or angular momentum removal due to MHD winds, is currently a subject of intense research \cite[e.g.,][]{Long2022, Trapman2023b}. Various diagnostic approaches have been proposed \citep[e.g.,][]{Alexander2023, Somigliana2023, Tabone2022}, including the examination of disc size evolution. Typically, viscous evolutionary models tend to predict disc spreading, whereas, in wind-dominated scenarios, disc sizes often remain relatively constant or may even shrink due to mass depletion. However, a central challenge lies in precisely defining disc size for meaningful comparisons with observed data \citep[e.g.,][]{Trapman2020, Toci2021, Toci2023}. Recent theoretical models have integrated MHD winds and determined that they can effectively explain both the dispersal and accretion characteristics of discs \citep{Zagaria2022, Tabone2022, Trapman2022}. Our analytical solution reveals that, in addition to factors like initial disc mass, accretion time-scale, and magnetic lever arm parameter, the initial disc size can profoundly influence disc evolution in an MHD wind-dominated regime. While investigating disc size evolution using our solutions to compare with existing observed PPDs, or undertaking population disc size synthesis, is beyond the scope of this study, our analytical solutions present a computational advantage for such studies.

When MHD winds are highly efficient, and the PPDs is substantial in size, the mechanisms governing planet formation within PPDs have more time to operate. Planet formation under these conditions presents an intriguing avenue for exploration, warranting the development of more sophisticated models. In addition to disc lifetime, the surface density gradually tends to adopt flatter profiles, influencing the radial drift of dust particles and even planet migration, which depends on the surface density slope. This phenomenon leads to the emergence of local gas pressure maxima due to wind launching, capable of trapping dust particles and enhancing their growth rates \citep{Taki2021, Arakawa2021, Zagaria2022}.

Our analytical solutions serve as a valuable tool for efficiently investigating the influence of wind strength and initial disc size on grain growth within the inner disc regions. Furthermore, previous research has shown that the emergence of MHD winds can inhibit Type I planet migration in the close-in region \citep{Ogihara2015, Ogihara2018} or even induce outward planet migration \citep{Kimmig2020}. Although our current study does not specifically address planet migration, our solutions retain relevance for conducting investigations in this domain.

In the context of MHD wind-dominated PPDs, our investigation yielded several key findings:

\begin{enumerate}
    \item The lifespan of these discs demonstrates a positive correlation with the magnetic lever arm, provided this parameter remains below approximately 6. Importantly, this correlation appears to be independent of the initial disc size.
    
    \item For PPDs with larger values of the magnetic lever arm \( \lambda \), the total disc mass remains notably higher than that of discs with smaller \( \lambda \) values, even after a substantial duration of around 20 characteristic timescales, equivalent to 20 million years, given our adoption of a characteristic timescale of about 1 million years.
    
    \item The ratio of mass ejection to accretion tends to increase over time in scenarios characterized by highly efficient winds, with this ratio surpassing unity. Notably, as the initial surface density profile becomes steeper, this ratio experiences a significant enhancement compared to discs that initiate their evolution with a flatter density profile.
    
    \item Furthermore, when examining the plane defined by the accretion rate at the inner disc edge and the total disc mass, we observed diverse trajectories contingent upon the values of the magnetic lever arm and the initial disc size.
\end{enumerate}

These findings offer valuable insights into the intricate dynamics governing MHD wind-dominated PPDs, shedding light on the role of various parameters in shaping their evolutionary trajectories.

\section*{Acknowledgements}

We express our gratitude to the referee for providing valuable insights through their insightful report, which greatly contributed to the improvement of this paper. 
All figures were generated with the \texttt{PYTHON}-based package \texttt{MATPLOTLIB} \citep{Hunter2007}.

%%%%%%%%%%%%%%%%%%%%%%%%%%%%%%%%%%%%%%%%%%%%%%%%%%
\section*{Data Availability}
The data underlying this article will be shared on reasonable request to the corresponding author.



\bibliographystyle{mnras}
\bibliography{example} % if your bibtex file is called example.bib

\begin{thebibliography}{}
\makeatletter
\relax
\def\mn@urlcharsother{\let\do\@makeother \do\$\do\&\do\#\do\^\do\_\do\%\do\~}
\def\mn@doi{\begingroup\mn@urlcharsother \@ifnextchar [ {\mn@doi@} {\mn@doi@[]}}
\def\mn@doi@[#1]#2{\def\@tempa{#1}\ifx\@tempa\@empty \href {http://dx.doi.org/#2} {doi:#2}\else \href {http://dx.doi.org/#2} {#1}\fi \endgroup}
\def\mn@eprint#1#2{\mn@eprint@#1:#2::\@nil}
\def\mn@eprint@arXiv#1{\href {http://arxiv.org/abs/#1} {{\tt arXiv:#1}}}
\def\mn@eprint@dblp#1{\href {http://dblp.uni-trier.de/rec/bibtex/#1.xml} {dblp:#1}}
\def\mn@eprint@#1:#2:#3:#4\@nil{\def\@tempa {#1}\def\@tempb {#2}\def\@tempc {#3}\ifx \@tempc \@empty \let \@tempc \@tempb \let \@tempb \@tempa \fi \ifx \@tempb \@empty \def\@tempb {arXiv}\fi \@ifundefined {mn@eprint@\@tempb}{\@tempb:\@tempc}{\expandafter \expandafter \csname mn@eprint@\@tempb\endcsname \expandafter{\@tempc}}}

\bibitem[\protect\citeauthoryear{{Alessi} \& {Pudritz}}{{Alessi} \& {Pudritz}}{2022}]{Alessi2022}
{Alessi} M.,  {Pudritz} R.~E.,  2022, \mn@doi [\mnras] {10.1093/mnras/stac1782}, \href {https://ui.adsabs.harvard.edu/abs/2022MNRAS.515.2548A} {515, 2548}

\bibitem[\protect\citeauthoryear{{Alexander}, {Rosotti}, {Armitage}, {Herczeg}, {Manara}  \& {Tabone}}{{Alexander} et~al.}{2023}]{Alexander2023}
{Alexander} R.,  {Rosotti} G.,  {Armitage} P.~J.,  {Herczeg} G.~J.,  {Manara} C.~F.,   {Tabone} B.,  2023, \mn@doi [\mnras] {10.1093/mnras/stad1983}, \href {https://ui.adsabs.harvard.edu/abs/2023MNRAS.524.3948A} {524, 3948}

\bibitem[\protect\citeauthoryear{{Anderson}, {Adams}  \& {Calvet}}{{Anderson} et~al.}{2013}]{Anderson2013}
{Anderson} K.~R.,  {Adams} F.~C.,   {Calvet} N.,  2013, \mn@doi [\apj] {10.1088/0004-637X/774/1/9}, \href {https://ui.adsabs.harvard.edu/abs/2013ApJ...774....9A} {774, 9}

\bibitem[\protect\citeauthoryear{{Andrews}}{{Andrews}}{2020}]{Andrews2020}
{Andrews} S.~M.,  2020, \mn@doi [\araa] {10.1146/annurev-astro-031220-010302}, \href {https://ui.adsabs.harvard.edu/abs/2020ARA&A..58..483A} {58, 483}

\bibitem[\protect\citeauthoryear{{Aoyama} \& {Bai}}{{Aoyama} \& {Bai}}{2023}]{Aoyama2023}
{Aoyama} Y.,  {Bai} X.-N.,  2023, \mn@doi [\apj] {10.3847/1538-4357/acb81f}, \href {https://ui.adsabs.harvard.edu/abs/2023ApJ...946....5A} {946, 5}

\bibitem[\protect\citeauthoryear{{Arakawa}, {Matsumoto}  \& {Honda}}{{Arakawa} et~al.}{2021}]{Arakawa2021}
{Arakawa} S.,  {Matsumoto} Y.,   {Honda} M.,  2021, \mn@doi [\apj] {10.3847/1538-4357/ac157e}, \href {https://ui.adsabs.harvard.edu/abs/2021ApJ...920...27A} {920, 27}

\bibitem[\protect\citeauthoryear{{Armitage}}{{Armitage}}{2011}]{Armitage2011}
{Armitage} P.~J.,  2011, \mn@doi [\araa] {10.1146/annurev-astro-081710-102521}, \href {https://ui.adsabs.harvard.edu/abs/2011ARA&A..49..195A} {49, 195}

\bibitem[\protect\citeauthoryear{{Armitage}, {Simon}  \& {Martin}}{{Armitage} et~al.}{2013}]{Armitage2013}
{Armitage} P.~J.,  {Simon} J.~B.,   {Martin} R.~G.,  2013, \mn@doi [\apjl] {10.1088/2041-8205/778/1/L14}, \href {https://ui.adsabs.harvard.edu/abs/2013ApJ...778L..14A} {778, L14}

\bibitem[\protect\citeauthoryear{{Bai} \& {Stone}}{{Bai} \& {Stone}}{2013}]{Bai2013}
{Bai} X.-N.,  {Stone} J.~M.,  2013, \mn@doi [\apj] {10.1088/0004-637X/769/1/76}, \href {https://ui.adsabs.harvard.edu/abs/2013ApJ...769...76B} {769, 76}

\bibitem[\protect\citeauthoryear{{Balbus} \& {Hawley}}{{Balbus} \& {Hawley}}{1991}]{Balbus1991}
{Balbus} S.~A.,  {Hawley} J.~F.,  1991, \mn@doi [\apj] {10.1086/170270}, \href {https://ui.adsabs.harvard.edu/abs/1991ApJ...376..214B} {376, 214}

\bibitem[\protect\citeauthoryear{{Blandford} \& {Payne}}{{Blandford} \& {Payne}}{1982}]{Blandford1982}
{Blandford} R.~D.,  {Payne} D.~G.,  1982, \mn@doi [\mnras] {10.1093/mnras/199.4.883}, \href {https://ui.adsabs.harvard.edu/abs/1982MNRAS.199..883B} {199, 883}

\bibitem[\protect\citeauthoryear{{Chambers}}{{Chambers}}{2019}]{Chambers2019}
{Chambers} J.,  2019, \mn@doi [\apj] {10.3847/1538-4357/ab2537}, \href {https://ui.adsabs.harvard.edu/abs/2019ApJ...879...98C} {879, 98}

\bibitem[\protect\citeauthoryear{{Ferreira}}{{Ferreira}}{1997}]{Ferreira1997}
{Ferreira} J.,  1997, \mn@doi [\aap] {10.48550/arXiv.astro-ph/9607057}, \href {https://ui.adsabs.harvard.edu/abs/1997A&A...319..340F} {319, 340}

\bibitem[\protect\citeauthoryear{{Ferreira} \& {Pelletier}}{{Ferreira} \& {Pelletier}}{1993}]{Ferreira1993}
{Ferreira} J.,  {Pelletier} G.,  1993, \aap, \href {https://ui.adsabs.harvard.edu/abs/1993A&A...276..625F} {276, 625}

\bibitem[\protect\citeauthoryear{{Hasegawa}, {Okuzumi}, {Flock}  \& {Turner}}{{Hasegawa} et~al.}{2017}]{Hasegawa2017}
{Hasegawa} Y.,  {Okuzumi} S.,  {Flock} M.,   {Turner} N.~J.,  2017, \mn@doi [\apj] {10.3847/1538-4357/aa7d55}, \href {https://ui.adsabs.harvard.edu/abs/2017ApJ...845...31H} {845, 31}

\bibitem[\protect\citeauthoryear{{Held} \& {Latter}}{{Held} \& {Latter}}{2018}]{Held2018}
{Held} L.~E.,  {Latter} H.~N.,  2018, \mn@doi [\mnras] {10.1093/mnras/sty2097}, \href {https://ui.adsabs.harvard.edu/abs/2018MNRAS.480.4797H} {480, 4797}

\bibitem[\protect\citeauthoryear{Hunter}{Hunter}{2007}]{Hunter2007}
Hunter J.~D.,  2007, Computing in Science \& Engineering, 9, 90

\bibitem[\protect\citeauthoryear{{Khajenabi}, {Shadmehri}, {Pessah}  \& {Martin}}{{Khajenabi} et~al.}{2018}]{Khajenabi2018}
{Khajenabi} F.,  {Shadmehri} M.,  {Pessah} M.~E.,   {Martin} R.~G.,  2018, \mn@doi [\mnras] {10.1093/mnras/sty153}, \href {https://ui.adsabs.harvard.edu/abs/2018MNRAS.475.5059K} {475, 5059}

\bibitem[\protect\citeauthoryear{{Kimmig}, {Dullemond}  \& {Kley}}{{Kimmig} et~al.}{2020}]{Kimmig2020}
{Kimmig} C.~N.,  {Dullemond} C.~P.,   {Kley} W.,  2020, \mn@doi [\aap] {10.1051/0004-6361/201936412}, \href {https://ui.adsabs.harvard.edu/abs/2020A&A...633A...4K} {633, A4}

\bibitem[\protect\citeauthoryear{{Kley}, {Papaloizou}  \& {Lin}}{{Kley} et~al.}{1993}]{Kley1993}
{Kley} W.,  {Papaloizou} J.~C.~B.,   {Lin} D.~N.~C.,  1993, \mn@doi [\apj] {10.1086/173268}, \href {https://ui.adsabs.harvard.edu/abs/1993ApJ...416..679K} {416, 679}

\bibitem[\protect\citeauthoryear{{Kratter} \& {Lodato}}{{Kratter} \& {Lodato}}{2016}]{Kratter2016}
{Kratter} K.,  {Lodato} G.,  2016, \mn@doi [\araa] {10.1146/annurev-astro-081915-023307}, \href {https://ui.adsabs.harvard.edu/abs/2016ARA&A..54..271K} {54, 271}

\bibitem[\protect\citeauthoryear{{Kunitomo}, {Suzuki}  \& {Inutsuka}}{{Kunitomo} et~al.}{2020}]{Kunitomo2020}
{Kunitomo} M.,  {Suzuki} T.~K.,   {Inutsuka} S.-i.,  2020, \mn@doi [\mnras] {10.1093/mnras/staa087}, \href {https://ui.adsabs.harvard.edu/abs/2020MNRAS.492.3849K} {492, 3849}

\bibitem[\protect\citeauthoryear{{Lodato} \& {Rice}}{{Lodato} \& {Rice}}{2004}]{Lodato2004}
{Lodato} G.,  {Rice} W.~K.~M.,  2004, \mn@doi [\mnras] {10.1111/j.1365-2966.2004.07811.x}, \href {https://ui.adsabs.harvard.edu/abs/2004MNRAS.351..630L} {351, 630}

\bibitem[\protect\citeauthoryear{{Lodato}, {Scardoni}, {Manara}  \& {Testi}}{{Lodato} et~al.}{2017}]{Lodato2017}
{Lodato} G.,  {Scardoni} C.~E.,  {Manara} C.~F.,   {Testi} L.,  2017, \mn@doi [\mnras] {10.1093/mnras/stx2273}, \href {https://ui.adsabs.harvard.edu/abs/2017MNRAS.472.4700L} {472, 4700}

\bibitem[\protect\citeauthoryear{{Long} et~al.,}{{Long} et~al.}{2022}]{Long2022}
{Long} F.,  et~al., 2022, \mn@doi [\apj] {10.3847/1538-4357/ac634e}, \href {https://ui.adsabs.harvard.edu/abs/2022ApJ...931....6L} {931, 6}

\bibitem[\protect\citeauthoryear{{Lynden-Bell} \& {Pringle}}{{Lynden-Bell} \& {Pringle}}{1974}]{Lynden-Bell1974}
{Lynden-Bell} D.,  {Pringle} J.~E.,  1974, \mn@doi [\mnras] {10.1093/mnras/168.3.603}, \href {https://ui.adsabs.harvard.edu/abs/1974MNRAS.168..603L} {168, 603}

\bibitem[\protect\citeauthoryear{{Najita} \& {Bergin}}{{Najita} \& {Bergin}}{2018}]{Najita2018}
{Najita} J.~R.,  {Bergin} E.~A.,  2018, \mn@doi [\apj] {10.3847/1538-4357/aad80c}, \href {https://ui.adsabs.harvard.edu/abs/2018ApJ...864..168N} {864, 168}

\bibitem[\protect\citeauthoryear{{Ogihara}, {Morbidelli}  \& {Guillot}}{{Ogihara} et~al.}{2015}]{Ogihara2015}
{Ogihara} M.,  {Morbidelli} A.,   {Guillot} T.,  2015, \mn@doi [\aap] {10.1051/0004-6361/201527117}, \href {https://ui.adsabs.harvard.edu/abs/2015A&A...584L...1O} {584, L1}

\bibitem[\protect\citeauthoryear{{Ogihara}, {Kokubo}, {Suzuki}  \& {Morbidelli}}{{Ogihara} et~al.}{2018}]{Ogihara2018}
{Ogihara} M.,  {Kokubo} E.,  {Suzuki} T.~K.,   {Morbidelli} A.,  2018, \mn@doi [\aap] {10.1051/0004-6361/201832720}, \href {https://ui.adsabs.harvard.edu/abs/2018A&A...615A..63O} {615, A63}

\bibitem[\protect\citeauthoryear{{Pascucci}, {Cabrit}, {Edwards}, {Gorti}, {Gressel}  \& {Suzuki}}{{Pascucci} et~al.}{2023}]{Pascucci2023}
{Pascucci} I.,  {Cabrit} S.,  {Edwards} S.,  {Gorti} U.,  {Gressel} O.,   {Suzuki} T.~K.,  2023, in {Inutsuka} S.,  {Aikawa} Y.,  {Muto} T.,  {Tomida} K.,   {Tamura} M.,  eds,  Astronomical Society of the Pacific Conference Series Vol. 534, Protostars and Planets VII. p.~567 (\mn@eprint {arXiv} {2203.10068}), \mn@doi{10.48550/arXiv.2203.10068}

\bibitem[\protect\citeauthoryear{{Rafikov}}{{Rafikov}}{2009}]{Rafikov2009}
{Rafikov} R.~R.,  2009, \mn@doi [\apj] {10.1088/0004-637X/704/1/281}, \href {https://ui.adsabs.harvard.edu/abs/2009ApJ...704..281R} {704, 281}

\bibitem[\protect\citeauthoryear{{Rafikov}}{{Rafikov}}{2017}]{Rafikov2017}
{Rafikov} R.~R.,  2017, \mn@doi [\apj] {10.3847/1538-4357/aa6249}, \href {https://ui.adsabs.harvard.edu/abs/2017ApJ...837..163R} {837, 163}

\bibitem[\protect\citeauthoryear{{Shadmehri} \& {Ghoreyshi}}{{Shadmehri} \& {Ghoreyshi}}{2019}]{Shadmehri2019}
{Shadmehri} M.,  {Ghoreyshi} S.~M.,  2019, \mn@doi [\mnras] {10.1093/mnras/stz2025}, \href {https://ui.adsabs.harvard.edu/abs/2019MNRAS.488.4623S} {488, 4623}

\bibitem[\protect\citeauthoryear{{Shakura} \& {Sunyaev}}{{Shakura} \& {Sunyaev}}{1973}]{Shakura1973}
{Shakura} N.~I.,  {Sunyaev} R.~A.,  1973, \aap, \href {https://ui.adsabs.harvard.edu/abs/1973A&A....24..337S} {24, 337}

\bibitem[\protect\citeauthoryear{{Shibaike} \& {Mori}}{{Shibaike} \& {Mori}}{2023}]{Shibaike2023}
{Shibaike} Y.,  {Mori} S.,  2023, \mn@doi [\mnras] {10.1093/mnras/stac3428}, \href {https://ui.adsabs.harvard.edu/abs/2023MNRAS.518.5444S} {518, 5444}

\bibitem[\protect\citeauthoryear{{Simon}, {Hughes}, {Flaherty}, {Bai}  \& {Armitage}}{{Simon} et~al.}{2015}]{Simon2015}
{Simon} J.~B.,  {Hughes} A.~M.,  {Flaherty} K.~M.,  {Bai} X.-N.,   {Armitage} P.~J.,  2015, \mn@doi [\apj] {10.1088/0004-637X/808/2/180}, \href {https://ui.adsabs.harvard.edu/abs/2015ApJ...808..180S} {808, 180}

\bibitem[\protect\citeauthoryear{{Somigliana}, {Testi}, {Rosotti}, {Toci}, {Lodato}, {Tabone}, {Manara}  \& {Tazzari}}{{Somigliana} et~al.}{2023}]{Somigliana2023}
{Somigliana} A.,  {Testi} L.,  {Rosotti} G.,  {Toci} C.,  {Lodato} G.,  {Tabone} B.,  {Manara} C.~F.,   {Tazzari} M.,  2023, \mn@doi [\apjl] {10.3847/2041-8213/acf048}, \href {https://ui.adsabs.harvard.edu/abs/2023ApJ...954L..13S} {954, L13}

\bibitem[\protect\citeauthoryear{{Suzuki}, {Muto}  \& {Inutsuka}}{{Suzuki} et~al.}{2010}]{Suzuki2010}
{Suzuki} T.~K.,  {Muto} T.,   {Inutsuka} S.-i.,  2010, \mn@doi [\apj] {10.1088/0004-637X/718/2/1289}, \href {https://ui.adsabs.harvard.edu/abs/2010ApJ...718.1289S} {718, 1289}

\bibitem[\protect\citeauthoryear{{Suzuki}, {Ogihara}, {Morbidelli}, {Crida}  \& {Guillot}}{{Suzuki} et~al.}{2016}]{Suzuki2016}
{Suzuki} T.~K.,  {Ogihara} M.,  {Morbidelli} A.,  {Crida} A.,   {Guillot} T.,  2016, \mn@doi [\aap] {10.1051/0004-6361/201628955}, \href {https://ui.adsabs.harvard.edu/abs/2016A&A...596A..74S} {596, A74}

\bibitem[\protect\citeauthoryear{{Tabone}, {Rosotti}, {Cridland}, {Armitage}  \& {Lodato}}{{Tabone} et~al.}{2022}]{Tabone2022}
{Tabone} B.,  {Rosotti} G.~P.,  {Cridland} A.~J.,  {Armitage} P.~J.,   {Lodato} G.,  2022, \mn@doi [\mnras] {10.1093/mnras/stab3442}, \href {https://ui.adsabs.harvard.edu/abs/2022MNRAS.512.2290T} {512, 2290}

\bibitem[\protect\citeauthoryear{{Takahashi} \& {Muto}}{{Takahashi} \& {Muto}}{2018}]{Takahashi2018}
{Takahashi} S.~Z.,  {Muto} T.,  2018, \mn@doi [\apj] {10.3847/1538-4357/aadda0}, \href {https://ui.adsabs.harvard.edu/abs/2018ApJ...865..102T} {865, 102}

\bibitem[\protect\citeauthoryear{{Taki}, {Kuwabara}, {Kobayashi}  \& {Suzuki}}{{Taki} et~al.}{2021}]{Taki2021}
{Taki} T.,  {Kuwabara} K.,  {Kobayashi} H.,   {Suzuki} T.~K.,  2021, \mn@doi [\apj] {10.3847/1538-4357/abd79f}, \href {https://ui.adsabs.harvard.edu/abs/2021ApJ...909...75T} {909, 75}

\bibitem[\protect\citeauthoryear{{Toci}, {Rosotti}, {Lodato}, {Testi}  \& {Trapman}}{{Toci} et~al.}{2021}]{Toci2021}
{Toci} C.,  {Rosotti} G.,  {Lodato} G.,  {Testi} L.,   {Trapman} L.,  2021, \mn@doi [\mnras] {10.1093/mnras/stab2112}, \href {https://ui.adsabs.harvard.edu/abs/2021MNRAS.507..818T} {507, 818}

\bibitem[\protect\citeauthoryear{{Toci}, {Lodato}, {Livio}, {Rosotti}  \& {Trapman}}{{Toci} et~al.}{2023}]{Toci2023}
{Toci} C.,  {Lodato} G.,  {Livio} F.~G.,  {Rosotti} G.,   {Trapman} L.,  2023, \mn@doi [\mnras] {10.1093/mnrasl/slac137}, \href {https://ui.adsabs.harvard.edu/abs/2023MNRAS.518L..69T} {518, L69}

\bibitem[\protect\citeauthoryear{{Trapman}, {Rosotti}, {Bosman}, {Hogerheijde}  \& {van Dishoeck}}{{Trapman} et~al.}{2020}]{Trapman2020}
{Trapman} L.,  {Rosotti} G.,  {Bosman} A.~D.,  {Hogerheijde} M.~R.,   {van Dishoeck} E.~F.,  2020, \mn@doi [\aap] {10.1051/0004-6361/202037673}, \href {https://ui.adsabs.harvard.edu/abs/2020A&A...640A...5T} {640, A5}

\bibitem[\protect\citeauthoryear{{Trapman}, {Tabone}, {Rosotti}  \& {Zhang}}{{Trapman} et~al.}{2022}]{Trapman2022}
{Trapman} L.,  {Tabone} B.,  {Rosotti} G.,   {Zhang} K.,  2022, \mn@doi [\apj] {10.3847/1538-4357/ac3ed5}, \href {https://ui.adsabs.harvard.edu/abs/2022ApJ...926...61T} {926, 61}

\bibitem[\protect\citeauthoryear{{Trapman}, {Rosotti}, {Zhang}  \& {Tabone}}{{Trapman} et~al.}{2023}]{Trapman2023b}
{Trapman} L.,  {Rosotti} G.,  {Zhang} K.,   {Tabone} B.,  2023, \mn@doi [\apj] {10.3847/1538-4357/ace7d1}, \href {https://ui.adsabs.harvard.edu/abs/2023ApJ...954...41T} {954, 41}

\bibitem[\protect\citeauthoryear{{Wang} \& {Goodman}}{{Wang} \& {Goodman}}{2017}]{Wang2017}
{Wang} L.,  {Goodman} J.~J.,  2017, \mn@doi [\apj] {10.3847/1538-4357/835/1/59}, \href {https://ui.adsabs.harvard.edu/abs/2017ApJ...835...59W} {835, 59}

\bibitem[\protect\citeauthoryear{{Weder}, {Mordasini}  \& {Emsenhuber}}{{Weder} et~al.}{2023}]{Weder2023}
{Weder} J.,  {Mordasini} C.,   {Emsenhuber} A.,  2023, \mn@doi [\aap] {10.1051/0004-6361/202243453}, \href {https://ui.adsabs.harvard.edu/abs/2023A&A...674A.165W} {674, A165}

\bibitem[\protect\citeauthoryear{{Wu}, {Chen}, {Jiang}, {Dong}, {Mac{\'\i}as}, {Lin}, {Rosotti}  \& {Elbakyan}}{{Wu} et~al.}{2023}]{Wu2023}
{Wu} Y.,  {Chen} Y.-X.,  {Jiang} H.,  {Dong} R.,  {Mac{\'\i}as} E.,  {Lin} M.-K.,  {Rosotti} G.~P.,   {Elbakyan} V.,  2023, \mn@doi [\mnras] {10.1093/mnras/stad1553}, \href {https://ui.adsabs.harvard.edu/abs/2023MNRAS.523.2630W} {523, 2630}

\bibitem[\protect\citeauthoryear{{Zagaria}, {Rosotti}, {Clarke}  \& {Tabone}}{{Zagaria} et~al.}{2022}]{Zagaria2022}
{Zagaria} F.,  {Rosotti} G.~P.,  {Clarke} C.~J.,   {Tabone} B.,  2022, \mn@doi [\mnras] {10.1093/mnras/stac1461}, \href {https://ui.adsabs.harvard.edu/abs/2022MNRAS.514.1088Z} {514, 1088}

\makeatother
\end{thebibliography}

%\begin{thebibliography}{99}
%\end{thebibliography}

%%%%%%%%%%%%%%%%%%%%%%%%%%%%%%%%%%%%%%%%%%%%%%%%%%

%%%%%%%%%%%%%%%%% APPENDICES %%%%%%%%%%%%%%%%%%%%%

%\appendix


%%%%%%%%%%%%%%%%%%%%%%%%%%%%%%%%%%%%%%%%%%%%%%%%%%

% Don't change these lines
\bsp	% typesetting comment
\label{lastpage}
\end{document}